\documentclass[twocolumn]{aastex63}
\usepackage{soulutf8} 

\def\fu{\mbox{4U~0142+61\ }}
\def\fuu{\mbox{4U~0142+61}}

\def\apjl{ApJ Lett.}
\def\aap{A\&A}

\def\fu{4U~0142+61\ }
\def\ffu{4U~0142+61}

\received{-}
\revised{-}
\accepted{-}
\submitjournal{ApJ}

\shorttitle{A \added{magnetic} white-dwarf \added{accretion} model for 4U~0142+61}
\shortauthors{Borges et al.}

\begin{document}

\title{\replaced{An}{A magnetic} white-dwarf accretion model for the anomalous X-ray pulsar 4U~0142+61}

\correspondingauthor{Sarah V. Borges}
\email{villanovaborges@gmail.com}

\author[0000-0003-4050-9920]{Sarah V. Borges}
\affiliation{Divis\~ao de Astrof\'\i sica -- Instituto Nacional de Pesquisas Espaciais/INPE \\
Av. dos Astronautas, 1758 \\ 
12227-010 -- S\~ao Jos\'e dos Campos/SP -- Brazil}

\author[0000-0002-9459-043X]{Claudia V. Rodrigues}
\affiliation{Divis\~ao de Astrof\'\i sica -- Instituto Nacional de Pesquisas Espaciais/INPE \\
Av. dos Astronautas, 1758 \\ 
12227-010 -- S\~ao Jos\'e dos Campos/SP -- Brazil}

\author[0000-0001-9386-1042]{Jaziel G. Coelho}
\affiliation{Divis\~ao de Astrof\'\i sica -- Instituto Nacional de Pesquisas Espaciais/INPE \\
Av. dos Astronautas, 1758 \\ 
12227-010 -- S\~ao Jos\'e dos Campos/SP -- Brazil}
\affiliation{Departamento de F\'isica, Universidade Tecnol\'ogica Federal do Paran\'a/UTFPR\\
85884-000 -- Medianeira/PR -- Brazil}

\author[0000-0002-4915-8672]{Manuel Malheiro}
\affiliation{Departamento de F\'isica, Instituto Tecnol\'ogico de Aeron\'autica/ITA \\
12228-900 -- S\~ao Jos\'e dos Campos/SP -- Brazil} 

\author[0000-0003-2162-8393]{Manuel Castro}
\affiliation{Divis\~ao de Astrof\'\i sica -- Instituto Nacional de Pesquisas Espaciais/INPE \\
Av. dos Astronautas, 1758 \\ 
12227-010 -- S\~ao Jos\'e dos Campos/SP -- Brazil}



\begin{abstract}
The quiescent emission of the anomalous X-ray pulsar~(AXP) 4U~0142+61 extends over a broad range of energy, from radio up to hard X-rays. In particular, this object is unique among soft gamma-ray repeaters~(SGRs) and AXPs in presenting simultaneously mid-\replaced{IR}{infrared} emission and pulsed optical emission. In spite of \replaced{having}{the} many propositions to explain this wide range of emission, it 
\replaced{is still lacking}{still lacks} one that reproduces \replaced{simultaneously}{all} the observations\deleted{in several wavelengths}. Filling this gap, we present a model 
\deleted{that is able}
to reproduce the quiescent spectral energy distribution \deleted{(SED)} of 4U~0142+61 from mid-infrared up to hard X-rays using plausible physical components and parameters. We propose that the persistent emission comes from \replaced{an}{a magnetic} accreting white dwarf~(WD) surrounded by a debris disk. This model assumes that: (i) the hard X-rays are due to the bremsstrahlung emission from the post-shock region of the accretion column; (ii) the soft X-rays are originated by hot spots on the WD surface; and (iii) the optical and infrared emissions are caused by an optically thick dusty disk, the WD photosphere, and the tail of the post-shock region emission. 
In this scenario, the fitted model parameters indicate that 4U~0142+61 harbors a fast-rotator magnetic near-Chandrasekhar WD, which is very hot and hence young. Such a WD can be the recent outcome of  a merger of two less massive WDs.
In this case, 4U~0142+61 can 
\replaced{be in a previous stage of a}{evolve to an} 
SN Ia and hence can give hints of the origin of these important astrophysical events. Additionally, we also present a new estimate of 4U~0142+61 distance, $3.78^{+0.12}_{-0.18}$~kpc, based on the measured Hydrogen column density and new interstellar extinction 3D maps.
\end{abstract}


\keywords{accretion, accretion disks --- stars: general --- stars: magnetic field --- stars: rotation --- stars: WDs --- X-rays: stars  --- stars: individual (4U~0142+61)}


\section{Introduction}
\label{sec:1}

Anomalous X-ray pulsars~(AXPs) \deleted{are objects that}
present spin period\added{s} of a few seconds \replaced{and characteristic}{,}
quiescent soft X-ray emission with a blackbody temperature of approximately 0.4~keV and luminosity of about $10^{33}$~erg~s$^{-1}$~\citep[e.g.,][]{2014ApJS..212....6O}. Initially,
AXPs were considered X-ray binaries.
However, due to the lack of a companion and a smaller soft X-ray temperature compared to X-ray binaries,
they were classified as a whole new group. Meanwhile, soft gamma-ray repeater\added{s}~(SGRs) 
\replaced{were discovered because of the}{are associated with} energetic outburst events. Presently, AXPs/SGRs are considered as 
\replaced{a}{the}
same class of objects, which are observationally characterized by a quiescent soft X-ray (2~--~10~keV) luminosity in the range $10^{30}$~--~$10^{35}$ erg~s$^{-1}$, period of 2~--~12 s, and spin-down of $10^{-15}$ to $10^{-10}$ s~s$^{-1}$~\cite[see][and references therein]{2014ApJS..212....6O}. In outburst, the energy can reach $10^{43}$~erg~\citep[see, e.g.,][]{2018MNRAS.474..961C}. Some AXPs/SGRs also present hard X-ray emission
\deleted{,}
as well as soft gamma-ray flares events. For comprehensive reviews on observations of AXP/SGRs, see \cite{2008A&ARv..15..225M}, \cite{2015RPPh...78k6901T}, and \cite{2017ARA&A..55..261K}.

The emission nature of AXPs/SGRs is still a reason for debate\replaced{ and}{. There are} several 
\added{proposed}
scenarios 
\deleted{have been proposed}
to explain their observed spectra and properties. The most accepted scenario is the magnetar model, which was first proposed by \citet{1992ApJ...392L...9D} and \citet{1995MNRAS.275..255T} and later developed by several other authors, such as \citet{2007ApJ...657..967B}, \citet{2012ApJ...748L..12R}, and \citet{2013ApJ...762...13B}. In this model, the AXP/SGRs present a huge magnetic field $(B)$, in the range 
\deleted{of}
$10^{13} -10^{15}$~G\replaced{, and their}{. Their}
persistent X-ray luminosity, as well as the bursts and flares typical of these sources~\citep{1979Natur.282..587M,1999Natur.397...41H,1999ApJ...510L.111H}, are believed to be powered by the decay of their ultra\added{-}strong magnetic fields. However, some limitations of the model, such as the discovery of the low-B ($<4.4\times10^{13}$~G) sources -- SGR~0418+5729, Swift~J1822.3--1606, and 3XMM~J185246.6+003317~\citep{2010Sci...330..944R,2012ApJ...754...27R,2013ApJ...770...65R,2011ApJ...743L..38L,2014ApJ...781L..16Z} -- have increased the interest for alternative scenarios. Some examples are the neutron star~(NS) accreting scenario, raised by \cite{1995A&A...299L..41V} and \cite{2001ApJ...554.1245A}, and the white dwarf~(WD) pulsar model~\citep{1990ApJ...365L...9P,malheiro+2012,2014PASJ...66...14C,2016IJMPD..2541025L}. 

4U~0142+61 is an AXP reported for the first time in {\it UHURU} catalog\deleted{ue} \citep{1983ARA&A..21...13B}. This object presents quiescent emission in a broad range of energy, from radio~\citep{2010ARep...54..995M} to hard X-rays \citep{2006ApJ...645..556K}\deleted{, in addition to some outburst and glitches events. This source is unique among AXPs/SGRs, since it presents mid-\replaced{IR}{infrared (IR)} emission and pulsed optical emission, which even separately are rare features in the class.}  Its period is 8.68~s and the spin-down is around $2.0 \times 10^{-12}$~s~s$^{-1}$ \citep{2014ApJS..212....6O}. The luminosities in soft and hard X-rays are estimated as $2.8\times10^{35}$ erg~s$^{-1}$ and $0.68\times10^{35}$ erg~s$^{-1}$, respectively, considering a distance of 3.6~kpc~\citep{2011PASJ...63..387E}.
\added{This source is unique among AXPs/SGRs, since it presents mid-infrared (IR) emission and pulsed optical emission, which even separately are rare features in the class. The outbursts and glitches in \fu are less energetic compared to the bulk of SGR/AXPs bursts \citep{2017ApJ...835...68G}. No flare has been observed in \fu so far~\citep{2014ApJS..212....6O}. }

In the magnetar paradigm, the IR emission of \fu
is interpreted as a passive fallback disk \citep{2006Natur.440..772W} and the hard and soft X-ray emissions were fitted by \cite{2014ApJ...786L...1H}\deleted{,} using a model based on large magnetic loops\added{, which was}
proposed by \cite{2013ApJ...762...13B}. Specifically, the hard X-rays emission is caused by the production of $e^-$-$e^+$ pairs close to the NSs surface\replaced{ and the}{. The} soft X-ray emission requires a combination of two modified black bodies \replaced{, one of them representing the emission of the photosphere and the other one, a hot spot}{from the NS photosphere and from a hot spot} \citep[see][for details]{2014ApJ...786L...1H}. The optical emission is reproduced by a power-law function, proposed to be of magnetospheric origin \citep{2006Natur.440..772W}.

In the accreting NS model, a debris disk is responsible for the IR and optical emissions \citep{2007ApJ...657..441E} and acts as a reservoir of matter for the accretion. The hard X-rays are caused by the accreting structure and the soft X-ray, by a polar cap in the NSs surface \citep{2013ApJ...764...49T, 2015MNRAS.454.3366Z}. 

In a third proposed scenario, that of a WD pulsar, the optical/IR data are explained by the WD photosphere and by a disk \citep{2013ApJ...772L..24R} and the X-ray emission is caused by a pulsar-like emission~\citep{malheiro+2012,2014PASJ...66...14C,2016IJMPD..2541025L}. However, no attempt to fit the X-ray emission \fu in this model was presented up to \replaced{date}{now}.

In this paper, we present a new scenario to explain the spectral energy distribution~(SED) of 4U~0142+61, from mid-\replaced{infrared}{IR} up to hard X-rays. We propose that the persistent emission comes from an accreting isolated magnetic WD surrounded by a debris disk, having gas and dusty regions.
This scenario is inspired by the periodic flux modulation and by the presence of mid-IR emission, which is rare for NSs. In fact, only three isolated NSs have detected mid-IR: the radio pulsars Crab, Vela, and Geminga \citep{2009A&A...504..525S,2011MNRAS.415..867D}. \replaced{And, among AXPs/SGRs,  only two have mid-IR emission, 1E~2259+586 \citep{2009ApJ...700..149K} and 4U~0142+61 \citep{2006Natur.440..772W}.} 
{Only two AXPs/SGRs have mid-IR emission: 1E~2259+586 \citep{2009ApJ...700..149K} and \fu \citep{2006Natur.440..772W}. }
Thus, mid-IR appears in about 0.3\% of all isolated NSs. On the other hand, the presence of mid-IR in WDs is quite common.  \cite{2011ApJS..197...38D} found that about 7\% of WDs presents an excess of mid-IR.

The presence of a dusty disk 
in \fu is corroborated by a 
possible emission feature around 9~$\mu m$, probably due to silicate~\citep{2008AIPC..983..274W}.
The presence of disks around isolated NSs is
still question for debate. \replaced{P}{A p}rotoplanetary disk is one possible origin of the planetary system around the PSR~B1257+12~\citep{1992Natur.355..145W,2001ApJ...550..863M}. \replaced{Also, f}{F}allback disks are \added{also} raised \deleted{as one possibility} to explain the observed braking index smaller than 3 \added{in some NSs} \citep{2001ApJ...554L..63M}. However, no isolated NS have confirmed debris disk. On the other hand, at least 27\% of \added{the} WDs presents traces of high elements that \replaced{only could}{could only} be explained by the accretion of material from planetary discs~\citep{2014A&A...566A..34K}. All these arguments reinforce a WD nature for \fuu.

This paper 
\added{presents a study of 4U~0142+61 emission in the context of a WD nature. It} is organized as follows. In Section \ref{sec:data}, we introduce the \fu observations that are used to test the proposed scenario. In Section \ref{sec:distance}, we estimate a new distance for \fu based on the 3D  reddening map of \cite{2018MNRAS.478..651G,2019arXiv190502734G} and compare it with previous estimates. In Section \ref{sec:model}, we present an overview of the model and its parameters. In Section \ref{sec:fit}, we show the spectral fit of 4U~0142+61 and discuss the derived parameters. In Section~\ref{sec:spin-down}, we derive the magnetic field of 4U~0142+61  from its spin-down. In Section~\ref{sec:origin_evol}, we discuss the probable origin and evolution of the object in our scenario. In Section~\ref{sec:glitches}, we discuss possible mechanisms for the glitches, bursts and radio emission of \fu in our model. Finally, in Section \ref{sec:conclusion}, we summarize our findings.

\section{The Spectral energy distribution of \mbox{4U~0142+61}}
\label{sec:data}

\explain{We added an initial "The" in the section title.}

\fu in quiescence emits in the X-rays, optical, mid/near-IR, and radio. In this \replaced{S}{s}ection, we describe these data  and \deleted{how we set up} the SED used in the modelling.

The \fu soft X-ray spectrum was observed using several telescopes in the past decades, e.g., \citet[][{\it ASCA}]{1996ApJ...463L..83W} and \citet[][{\it Chandra}]{2002ApJ...568L..31J}. In our fit, we used the data \replaced{reduced by}{from} \citet[][{\it Suzaku}]{2010ApJ...722L.162E}\footnote{Observations taken in 2007-08-13 04:04:13 (seq. number 402013010). Kindly provided by the authors.}, which is deconvolved from the instrumental response. 
The hard X-ray emission was discovered by \citet[][ \textit{INTEGRAL, 20~--~300~keV}]{2006ApJ...645..556K} \deleted{in the 20~--~300~keV range} and was also observed by other telescopes, such as {\textit NuSTAR} 
in 0.5~--~79~keV \citep{2015ApJ...808...32T} and {\textit Suzaku} in 15~--~60~keV~\citep{2017ApJS..231....8E}. In our fit, we use\added{d the} {\it INTEGRAL} data, whose reduction is presented in Sec.~\ref{sec:integral}

The optical emission of 4U~0142+61 was discovered by \cite{2000Natur.408..689H}. Complementary photometry was performed by \cite{2004A&A...416.1037H}, \cite{2005MNRAS.363..609D}, and \cite{2016MNRAS.458L.114M}, who also present the first optical spectrum of 4U~0142+61. 
\replaced{ picked out}{We used} the {\it Gran Telescopio Canarias }~(GTC) optical data from \citet[][see Table \ref{tab:dados_opt}]{2016MNRAS.458L.114M}\replaced{, because they comprise a quasi-simultaneous data set in optical bands 
have a homogeneous reduction and a careful differential photometry calibration.}
{This quasi-simultaneous data set has been homogeneously reduced and a careful differential photometric calibration was performed.} This data set is consistent with previous measurements. 

The near-IR \textit{K}-band flux was measured for the first time in 1999 using {\it Keck} telescope by \citet{2004A&A...416.1037H}. After that, several near-IR observations were performed. Even though most of the them are in the \textit{K}-band, data in \textit{J} and \textit{H} bands have been taken using {\it CFHT}, {\it Subaru}, and {\it Gemini}~\citep{2006ApJ...652..576D}.  For the near-IR, we selected the \textit{Gemini} \textit{J}, \textit{H}, and \textit{ K$_{s}$} observations from \citet[][see Table 
\ref{tab:dados_opt}]{2006ApJ...652..576D}, because they provide a smoother SED when combined with the GTC optical data and the mid-IR {\it Spitzer} in 4.5 and 8.0~$\rm{\mu}$m  data from \cite{2006Natur.440..772W}, 
\replaced{as we discuss below or in Sec.~}{which are also used to model the \fu SED.}\ref{sec:ajuste}.

\fu emission is strongly affected by interstellar absorption and extinction, hence the data must be corrected by this effect. 
We adopt\added{ed} an interstellar Hydrogen column density, $N_H$, of $6.4 \times 10^{21}$~cm$^{-2}$\added{. This value was} obtained using individual absorption edges of metals in X-ray spectrum \citep{2006ApJ...650.1082D}.
The X-rays absorption cross-section are from \cite{1983ApJ...270..119M}, who presented an approximate analytic expression. 
To calculate the optical and near-IR extinction, we consider\added{ed} the 
\added{ratio of total to selective} extinction \deleted{coefficient} $R_V = 3.1$ and the relation between $N_H$ and the extinction \textit{A}$_\lambda$ from \cite{2017MNRAS.471.3494Z}\replaced{to}{. Hence, we} obtain\added{ed}

\begin{equation}\label{eq:NH}
    \frac{A_\lambda}{R_\lambda} = (1.55\pm0.015)\times10^{-22}N_H.
\end{equation}

We use\added{d} the extinction  \replaced{coefficient}{law} $R_\lambda$ from \cite{1989ApJ...345..245C}. The \added{obtained} values of extinction for \added{\fu in} each band are shown in Table \ref{tab:dados_opt}.

\begin{table}[ht]
\begin{center}
\caption{Optical and near-IR data and reddening used in the \fu modelling
}
\label{tab:dados_opt}
\begin{tabular}{ccccc}
\hline
\hline
Band & \textit{A}$_\lambda$ & Observed Mag. &Date & Ref. \\
\hline
\textit{g} &3.67  & 27.37 $\pm$ 0.25(0.58) & 2013 Aug 09 & 1 \\
\textit{r} &2.67 & 25.79 $\pm$ 0.07(0.26) & 2013 Aug 09 & 1 \\
\textit{i} & 2.01& 24.55 $\pm$ 0.05(0.22) & 2013 Aug 09 & 1 \\
\textit{z} &1.44 & 23.76 $\pm$ 0.07(0.28) & 2013 Aug 09 & 1 \\
\textit{J} &0.87 & 21.97 $\pm$ 0.16 & 2004 Nov 02 & 2 \\
\textit{H} & 0.59& 20.69 $\pm$ 0.12 & 2004 Nov 02 & 2 \\
\textit{$K_s$} &0.36 & 19.96 $\pm$ 0.07 & 2004 Nov 02 & 2 \\
\hline
\end{tabular}
\end{center}
\textbf{Notes.} The errors in parenthesis represent the zero-point error and were also considered in the fitting procedure.
 
\textbf{References.} (1)~\cite{2016MNRAS.458L.114M}; (2)~\cite{2006ApJ...652..576D}.
\end{table}

\begin{figure*}[tbhp]
\includegraphics[clip, trim=1.5cm 9.5cm 2cm 1cm]{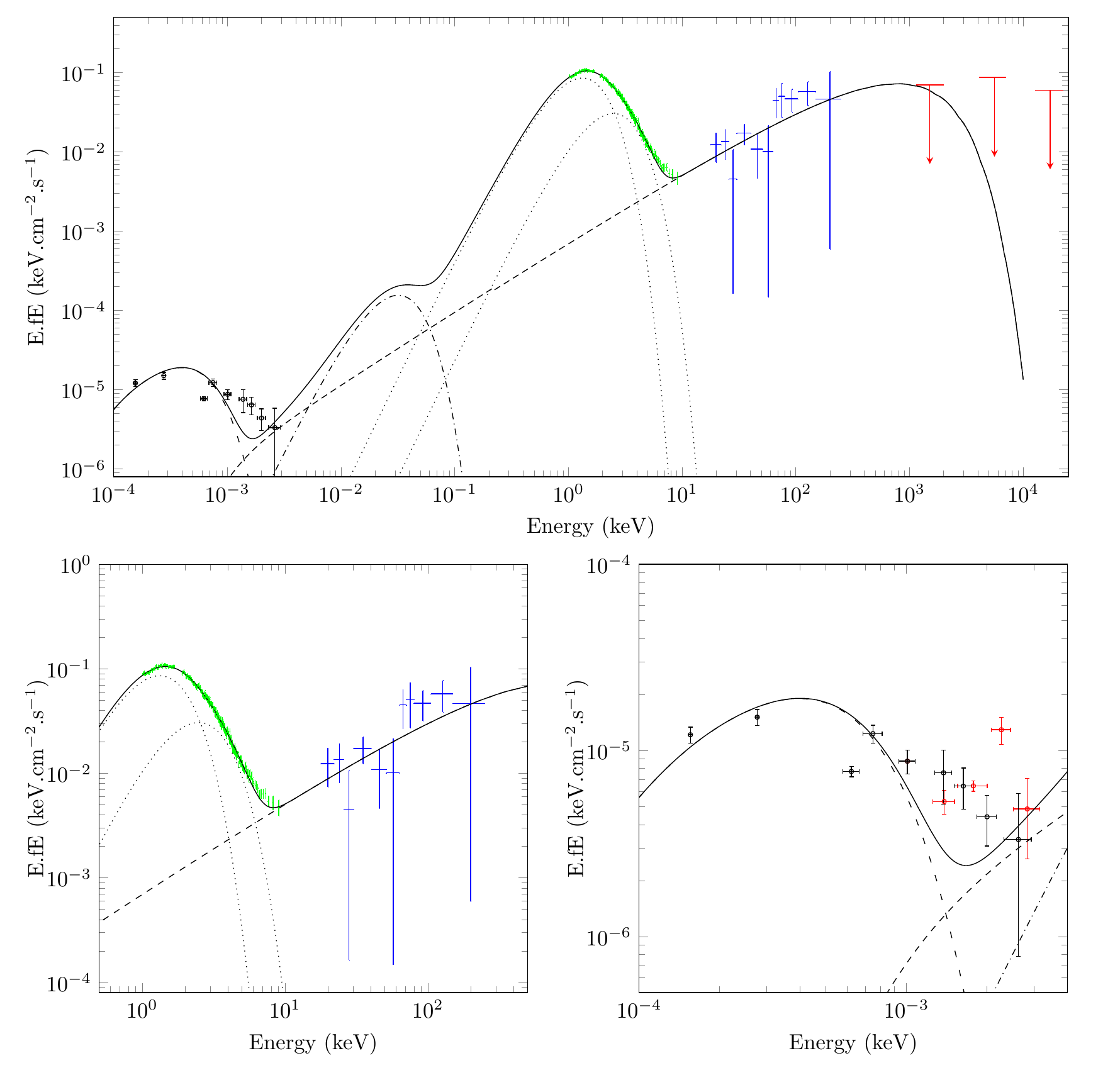}
\caption{\explain{We have replaced the bremsstrahlung component from the optically thin approximation to the correct radiative transport solution, which presents the break at low frequencies.} Derreddened and \replaced{deabsorved}{deabsorbed} SED of 4U~10142+61 along with the best fit.  The continuous black curve is the complete fit, the long-dashed curve is the disk component, the dot-dashed curve is the WD photosphere, the two dotted curves are the hot spots components, and the short-dashed curve is the bremsstrahlung component. 
The black crosses are from \citet[mid-infrared, {\it Spitzer}]{2006Natur.440..772W}, \citet[near-infrared,{\it Gemini}]{2006ApJ...652..576D} and \citet[optical, {\it GTC}]{2016MNRAS.458L.114M}; green crosses \replaced{are}{represent the} soft X-ray data from \citet[{\it Suzaku}]{2010ApJ...722L.162E}; \added{the} blue crosses are the \deleted{combined reduced data from} {\it INTEGRAL} \added{data} (see Sec.~\ref{sec:integral}), and the orange upper limits in gamma-rays are from \citet[ {\it COMPTEL}]{denhartog2008}.
{\it Top panel:} The entire spectral range, from mid-infrared up to gamma-rays. {\it Left-bottom panel:} Zoom at the high-energ\replaced{ies}{y} end. {\it Right-bottom panel:} Optical and infrared region. \added{The red points represent the measurements from \cite{2000Natur.408..689H,2004A&A...416.1037H}.}}
\label{fig:SED}
\end{figure*}

The dereddened and \replaced{deabsorved}{deaborbed} \fu SED adopted to fit our model is shown in Figure~\ref{fig:SED}. This figure also displays 
\added{the optical points from \citet{2000Natur.408..689H,2004A&A...416.1037H} and}
the upper limit for gamma-ray fluxes from \citet[][\textit{{\it CGRO COMPTEL}}]{denhartog2008}. 
\added{The} \fu \deleted{has a proposed} radio emission at  111~MHz \replaced{with a measured flux density of $S=$}{is} $30\pm20$~mJy \citep{2010ARep...54..995M}, which we \replaced{do}{did} not attempt to fit in this paper.

\subsection{INTEGRAL data reduction}
\label{sec:integral}

In our \fu modelling, we use\added{d} the 2007 data available on the {\it INTEGRAL} database and presented in Table~\ref{tab:integral}.
These data were first presented  by~\cite{denhartog2008}.
We searched for data within a 14.5$^{\circ}$-wide field of view and a minimum exposure time ($good\_isgri$) of 100\,s. To reduce and obtain the spectrum, we used the OSA 10.2 software and the recipes to the IBIS/ISGRI telescope \citep{2003A&A...411L.141L,2003A&A...411L.131U}. In each task, default parameter values were used.
The data in the entire revolution group were used to obtain one spectrum. 
To deconvolve this spectrum from instrumental effects, 
we fitted it in the 20 -- 200\,keV energy range with a power-law component in order to obtain the unfolded spectrum.
This fit was performed using the XSPEC software~\citep{1996ASPC..101...17A}. The resulting unfolded spectrum is used in our model fitting. 

\begin{table*}
\centering
\caption{INTEGRAL observations.}
\label{tab:integral}
\begin{tabular}{ccccc}
\hline \hline
Revolutions		&  MJD	& Time span	& ScWs			&$t_{eff}$(ks)	 \\
\hline
528 ToO 		& 54139.7661 --  54142.0288 	&	2007 Feb 08~--~2007 Feb 11	&  	50	&114	\\		
\hline
\end{tabular}
\\
 \textbf{Notes:} ScWs is the number of Science Windows used to obtain the spectra and $t_{eff}$ is the effective observation time.
\end{table*}

\section{Distance of 4U~0142+61}
\label{sec:distance}

Our aim is to present a model that fits not only the shape of \fu SED but also the level of the observed emission from mid-IR to X-rays. Hence\added{,} we refine\added{d the} \fu distance as presented below.

\cite{2006ApJ...650.1070D} estimate\added{d} the distance of \fu using the following methodology.  They  selected the red clump stars, \deleted{an} \replaced{H}{h}elium-core burning giant\added{s}, \replaced{in}{from} the {\it 2MASS}~catalog along the line of sight of \fu and derived \replaced{how the extinction varies}{the variation of extinction} according to the distance. Thus, they estimated the distance of $3.6\pm0.4$~kpc using $N_H = 6.4 \times 10^{21}$~cm$^{-2}$ and the relation $N_H/A_V=(1.79 \pm 0.03)\times 10^{21}$
cm$^{-2}$\, mag$^{-1}$
\citep[see][]{1995A&A...293..889P}. This method presents some limitations such as the small amount of red clump stars and the contamination by other types of stars, once the identification of red clump giants was done solely by their position in the color-magnitude diagram. 

In an attempt to get a more accurate distance, we used the updated relation of $N_H/A_V=(2.08 \pm 0.02)\times 10^{21}$ cm$^{-2}$\,mag$^{-1}$
from \cite{2017MNRAS.471.3494Z}. The resulting difference in the value of $A_V$, and hence E(B~-~V),
is approximately 15\%. Moreover, we use\added{d} the online version\footnote{\url{http://argonaut.skymaps.info/}} of the Galactic reddening 3D map from \cite{2018MNRAS.478..651G,2019arXiv190502734G}, which associates the distance with $E(B-V)=A_V/R_V$, to get a new estimate of \added{the} distance for 4U 0142+61. Using the relation from \cite{2017MNRAS.471.3494Z} and $R_V=2.742$ (see \citealt{2018MNRAS.478..651G,2019arXiv190502734G} and Table 6 of \citealt{2011ApJ...737..103S}), we \replaced{find}{found} $E(B-V)=1.12\pm0.04$~mag. Thus, our distance estimate\deleted{d} for \fu is $3.78^{+0.12}_{-0.18}$~kpc, which is used throughout this paper.

\section{A magnetic WD accretion model} 
\label{sec:model}

\explain{We have changed the section title. The original title was "An accretion model for isolated WDs".}

In this section, we describe the proposed scenario and the quantitative model for the emission of 4U~0142+61 as an isolated WD. We propose that the persistent emission comes from the WD photosphere, a disk, and \replaced{an}{a magnetic} accretion column. This scenario is inspired by: (i) the periodic flux modulation, which could be explained by a accretion column; and (ii) the detected IR emission and silicate line emission, which indicate\deleted{s} the presence of a disk.

\replaced{The disk is formed by an gaseous internal region and its external part is composed of dust and gas.}{The disk has a gaseous inner region and an outer portion made up of dust and gas.}
In spite of \replaced{having also}{also having} gas, we call the external portion of dusty disk.
Even though only the dusty disk is observed in most WDs, the presence of the gaseous region has already been confirmed for some objects~\citep{2006Sci...314.1908G,2007MNRAS.380L..35G,2008MNRAS.391L.103G,2012ApJ...751L...4M}. 

The dusty disk is optically thick and emits such as a multi-temperature blackbody. The temperature of its inner radius is the grain sublimation temperature, which is about 1\,500~K for silicates. Conversely, the gaseous disk is optically thin and its emission can be neglected \added{(see Sec.~\ref{sec:depth})}. \deleted{In Sec.~\ref{sec:depth}, we discuss if this supposition is valid.} The inner radius of the gaseous disk is equal to the magnetosphere radius. For that point on, the matter flows into the WD surface following the magnetic field lines and the debris disk ceases to exist. 

Close to the WD photosphere, the in-falling flow of matter \replaced{suffers a shock}{suffers a collisional shock}, forming an extremely hot region, the so called post-shock region that emits bremsstrahlung \cite[see, e.g.,][]{2017PASP..129f2001M}. About half of that energy reaches the WD surface, where it is absorbed and re-emitted in lower energies,
forming hot spots. Once the high-energy emission of 4U~0142+61 is pulsed, with two peaks per phase, 
it is plausible that we see the emission from two different accreting regions.

To sum up, the emitted flux of \fu can be expressed by

\begin{equation}
\label{eq:flux_total}
F_{total} = F_{disk} + F_{WD} + F_{spot} + F_{brem}.
\end{equation}

Next sections are devoted to explain each term of equation above, expanding all of them in their parameters. First, we discuss the blackbody flux emitted by the photosphere of the WD ($F_{WD}$) and the hot spot ($F_{spot}$). \replaced{After}{Then}, we comment on the emission from the accretion column ($F_{brem}$). At last, we describe the emission from the disk ($F_{disk}$). 

\subsection{WD photosphere and hot spots}

We assume\added{d} that the WD photosphere emits
such as \added{a} blackbody, in which the intensity for a given wavelength $\lambda$ and temperature $T$ is the Planck function, $B(\lambda,T)$. The\added{refore, the} photosphere flux~($F_{WD}$) \replaced{was calculated by}{is}

\begin{equation} \label{eq:flux_photosphere}
    F_{WD}(\lambda, T_{WD},R_{WD},d) =\pi B(\lambda,T_{WD})
    \bigg(\frac{R_{WD}}{d}\bigg)^2, 
\end{equation}

\noindent 
where $T_{WD}$ is the effective temperature of the photosphere and $R_{WD}$  is the radius of the WD. $F_{WD}$ is fitted simultaneously with $F_{disk}$ and after the accretion column and hot spot\added{s} (see Section \ref{sec:ajuste}). 
\replaced{This way, $R_{WD}$, which is estimated by the hard X-rays fitting, 
is a fixed parameter for the optical fitting, whereas $T_{WD}$ is a free parameter.}
{Hence, $T_{WD}$ is a free parameter, but $R_{WD}$ is a fixed parameter in the optical and IR fitting, because it is previously estimated by the hard X-rays fitting.}

The flux for each hot spot on the WD surface is also assumed as a blackbody and is given by

\begin{equation} \label{eq:flux_spot}
    F_{spot}(\lambda, T_{spot},R_{spot},d) =\pi B(\lambda,T_{spot}) \bigg(\frac{R_{spot}}{d}\bigg)^2,
\end{equation}

\noindent where $T_{spot}$ is the temperature of the spot and $R_{spot}$ is the radius of the spot. $F_{spot}$ is the main emission mechanism in the soft X-rays. In \fu fit, we consider\added{ed} two spots, whose parameters are discriminated by indexes 1 and 2.

\subsection{Post-shock region}

We assume\added{d} that the post-shock region emits by thermal bremsstrahlung. According to \cite{1986A&AS...65..511M}, the bremsstrahlung emitted power is

\begin{eqnarray}
\label{eq:emissivity}
P(\lambda,T_{brem}) = 2.051 \times 10^{-22} A\, g_{ff} n_e^2 \lambda^{-2}\times\\ \nonumber \times T_{brem}^{-1/2} exp\Bigg(\frac{-143.9}{\lambda T_{brem}}\Bigg).
\end{eqnarray}

\explain{We have included the factor A in the equation above. It was considered in the calculations, but it was not explicit in the equation.}

The parameters $n_e$, $T_{brem}$,
and $g_{ff}$ are the electron number density, the temperature of the bremsstrahlung emission, and the Gaunt factor, respectively. For the range of temperatures found in \replaced{accretions columns}{post-shock regions of WDs}, it is necessary to apply the relativistic correction, $A$, found in \cite{rybicki1979radiative}

\begin{eqnarray}
\label{eq:relativistic}
A = \left\{ 
\begin{array}{ll}
1 & \textnormal{  , for $T_{brem}\leq9.1$~keV};\\ 
1+4.4\times10^{-10}T & \textnormal{  , for $T_{brem}>9.1$~keV}.
\end{array}
\right.
\end{eqnarray}

For temperatures below $\sim100$~keV, we can consider a non-relativistic $g_{ff}$~\citep{1961ApJS....6..167K,1998MNRAS.300..321S}. However, for higher temperatures, which is the case of our fit, as we show later, we must take into account the relativistic $g_{ff}$. Thus, we adopt\added{ed} $g_{ff}$ from \cite{1998ApJ...507..530N}, \cite{2000ApJS..128..125I}, and \cite{2015MNRAS.449.2112V}. To estimate $g_{ff}$, it is also necessary to evaluate the ion charge of the in-falling matter, which depends on its chemical composition. Hence\added{,} \replaced{we need}{it was necessary} to define the disk material. \fu in the WD accreting model is probably the product of the merger of two CO WDs (see Sec.~\ref{sec:origin_evol})\replaced{ and hence}{. Therefore,} we adopt\added{ed} a disk composition \replaced{mainly by}{of} carbon~\citep{2009A&A...500.1193L}.

\added{Assuming that the region is cylindrical, with a height $H_{brem}$, the optical depth of the bremsstrahlung emission, $\tau_{brem}$, is:}

\begin{equation}\label{eq:depth_brem}
    \tau_{brem}=\frac{H_{brem}P(\lambda, T_{brem})}{4\pi B(\lambda, T_{brem})}.
\end{equation}

Assuming that the 
\deleted{region is optically thin and cylindrical, with a height $H_{brem}$ and} radius \added{is} $R_{brem}$, the flux of the bremsstrahlung emission can be written as


\begin{eqnarray}\label{eq:brem}
 \nonumber F_{brem}(n_e,R_{brem},H_{brem},T_{brem},d) = \\
    =(1-e^{-\tau_{brem}})B(\lambda, T_{brem})\pi\bigg(\frac{R_{brem}}{d}\bigg)^2.
\end{eqnarray}

\explain{The equation above was modified from the optically thin approximation to the version with no assumption relative to the optical depth.}

It is useful to define a normalization factor, $N$, related to the total squared number density of emitting particles in the region\replaced{.}{:} 

\begin{equation}\label{eq:emission}
   N =  \int n_e^2 dV.
\end{equation}

The value of $n_e$, in cm$^{-3}$, of the post-shock region can be calculated \replaced{by}{as} \citep[see][]{2002apa..book.....F}

\begin{eqnarray}\label{eq:ne}
  \nonumber n_e = 5.9 \times 10^{14} \bigg(\frac{\dot M}{10^{16}\text{~g~s}^{-1}}\bigg)\bigg(\frac{M_{WD}}{M_{\odot}}\bigg)^{-1/2}\times\\
   \times\bigg(\frac{R_{WD}}{10^9\text{~cm}}\bigg)^{-3/2}\bigg(\frac{f}{10^{-2}}\bigg)^{-1}\text{~cm}^{-3},
\end{eqnarray}

\noindent which depends on $M_{WD}$, $R_{WD}$, the accretion rate $\dot M$, and the fraction {\it f} of the WD photosphere \replaced{occupied by accretion.}{area covered by the footprint of the accretion column.}

The upper limit of $H_{brem}$ is given by~\citep{2002apa..book.....F}

\begin{eqnarray}
\nonumber H_{brem} < 9\times10^8\bigg(\frac{\dot M}{10^{16}\text{~g~s}^{-1}}\bigg)^{-1}\bigg(\frac{M_{WD}}{M_{\odot}}\bigg)^{3/2}\times \\
\times\bigg(\frac{R_{WD}}{10^9\text{~cm}}\bigg)^{1/2}\bigg(\frac{f}{10^{-2}}\bigg)\text{~cm}.
\label{eq:height}
\end{eqnarray}

We adopt\added{ed} the following estimate for $\dot M$\deleted{,}

\begin{equation}\label{eq:mdot}
\dot M = \frac{L_{brem} R_{WD}}{GM_{WD}},  
\end{equation}

\noindent where $G$ is the gravitational constant and $L_{brem}$ is the accretion luminosity of the model and is related to \added{the bremsstrahlung flux,} equation~\ref {eq:brem}.

\added{The emission from the post-shock region is optically thin in the X-rays range, so it is proportional to the volume of the emitting region, $V = H R_{brem}^2$.}
We \added{artificially} consider\added{ed} $R_{brem}=\sqrt{(R_{spot 1}^2+R_{spot 2}^2)}$ \deleted{in order} to guarantee that the accretion column and hot spots \replaced{occupy}{have} similar areas. \deleted{Moreover, $f$ is $R_{brem}^2/R_{WD}^2$.}

An analytic formula for the post-shock temperature of a WD ($T_{shock}$) was firstly derived by \cite{Aizu}. Since then, improvements on this formula were presented by several authors. In this paper, we used the expression from \citet{2016temp},

\begin{equation}\label{tempnovo}
   T_{shock} = \frac{3GM_{WD} m_H\mu}{8kR_{WD}}\left(1-\frac{R_{WD}}{R_{m}}\right).
\end{equation}

\added{\noindent  where $m_H$ is the mass of hydrogen atom, $k$ is the Boltzmann constant, $R_{m}$ is the magnetospheric radius, and $\mu$ is the mean molecular weight of the in-falling gas, which we considered 1.714 for a carbon disk.}

We adopt\added{ed} the \added{following} relation between $T_{brem}$ and $T_{shock}$ \citep[also from][]{2016temp},

\begin{equation}\label{eq:temp_brem}
kT_{brem}=0.64kT_{shock}.    
\end{equation}

\deleted{In equation \ref{tempnovo}, $m_H$ is the mass of hydrogen atom, $k$ is the Boltzmann constant, $\mu$ is the mean molecular weight of the in-falling gas, which we consider 1.714 for the disk composition, and $R_{m}$ is the magnetospheric radius.}

To estimate $T_{shock}$ and $T_{brem}$, we should adopt a value for \added{the} magnetospheric radius $R_{m}$. We consider\added{ed} $R_m$ equal to the corotational radius\added{, $R_{c}$,}

\begin{equation}\label{eq:corotational}
    R_{c} = \bigg(\frac{GM_{WD}}{\omega ^2}\bigg)^{1/3},
\end{equation}

\noindent where $\omega$\added{=$2\pi/T$} is the angular speed\deleted{ $2\pi/T$}, being $T$ the WD rotation period. In Section \ref{sec:spin-down}, we discuss the relation between $R_m$ and $R_c$ and confirm the validity of this assumption.

To sum up, $F_{brem}$ can be represented by

\begin{eqnarray} \label{eq:brem_emission}
\nonumber F_{brem}(N,T_{brem},d) = \\ =F_{brem}(n_e,R_{brem},H_{brem},T_{brem},d).
\end{eqnarray}

Therefore, the hard X-rays fit can be performed considering only $N$ and $T_{brem}$. All \deleted{the} other parameters~($M_{WD}$, $R_{WD}$, $\dot M$, $L_{brem}$, $n_e$, $H_{brem}$, $R_{brem}$, $f$) are derived from $N$, $T_{brem}$, $R_{spot 1}$, and $R_{spot 2}$.

\subsection{Multi-temperature disk}
\label{sec:multi_temp_disk}

Disks around WDs are common for both young and old populations~\citep{2014A&A...566A..34K}.
For instance, several old cool metal-rich WDs have small disks around them, supposedly caused by tidal disruption of small bodies \citep{2003ApJ...584L..91J}. These disks are optically thick and geometrically thin \citep{2012MNRAS.423..505M} and usually emit from mid-IR to optical wavelengths. The inner and outer radii \replaced{are around}{range from} a fraction \added{of} to a few $R_{\odot}$. In contrast, young hot WDs, such as those of the Helix planetary nebula, have large disks, which are probably created by collision of Kuiper-Belt-like objects~\citep{2011AJ....142...75C}. Those disks emit in the far to mid-IR and the radii are approximately a few AU \citep{2011AJ....142...75C}. In both cases, the WD accretes matter from the disk.

\replaced{Once}{As} \fu emits from mid-IR to optical, we fit\added{ted} the data adopting the disk flux relation proposed by \cite{1997ApJ...490..368C} for disks around T Tauri stars. They are passive disks, which reprocess the emission from star and accretion column. Hence\added{,} the dominant heating mechanism is irradiation.  This same model was used by
\cite{2003ApJ...584L..91J} to fit the disk of the cold WD \mbox{G29-38} and by \cite{2013ApJ...772L..24R} to fit the IR emission of \fu in the WD pulsar scenario. According to \citet{1997ApJ...490..368C} and \citet{2003ApJ...584L..91J},
the emitted flux from the debris disk in a given frequency $\nu$ can be expressed by

\begin{eqnarray}
\label{eq:flux_disk}
\nonumber  F_{disk}(\nu, T_{in},T_{out},T_{WD}, R_{WD}, d, \dot M) & 
= 12\pi^{1/3}\cos(i)\times\\ 
\nonumber \Bigg(\frac{R_{WD}}{d}\Bigg)^2\Bigg(\frac{2kT_{tot}}{3h\nu}\Bigg)^{8/3} \times & \\
\Bigg(\frac{h\nu^3}{c^2}\Bigg)\int^{x_{out}}_{x_{in}}\frac{x^{5/3}}{e^x-1}dx, &
\end{eqnarray}

\noindent where\deleted{,} {\it i} is the inclination of the disk, {\it h} is the Planck constant, c is the speed of light, $x=h\nu/kT_{irr}(R)$ and, $T_{irr}(R)$ is the debris disk temperature, which ranges from $T_{out}$ to $T_{in}$. $T_{tot}$ quantifies the disk heating by the two heating sources in the system, namely the WD and the accretion column, and is given by \citep{1997ApJ...490..368C}:

\begin{equation}
T_{tot}=(1-A_d)^{3/16}\left[T_{WD}+\left(\frac{G\dot M M_{WD}}{ \sigma R_{WD}}\right)^{0.25}\right],
\end{equation}

\noindent where $\sigma$ is the Stefan-Boltzmann constant and  $A_d$ is the albedo. It is important to include the effect of the albedo\deleted{,} because part of incident flux on the disk is reflected and does not contribute to the heating.

It is assumed that the temperature decays as $R^{-3/4}$ (see equation \ref{eq:radius_temp}). Thus, each $T_{irr}(R)$ value has an associated radius, $R$, in such a way that a given value of $T_{in}$ ($T_{out}$) corresponds to a value of  $R_{in}$ ($R_{out}$).


\begin{equation}
\label{eq:radius_temp}
    T_{irr}(R)=T_{tot}\left(\frac{R}{R_{WD}}\right)^{-3/4}.
\end{equation}

\explain{The subscript of T was corrected as noted by the referee.}

For radii between $R_m$ and $R_{in}$, the disk is gaseous and does not emit (see Sec. \ref{sec:depth}).
In addition, we consider\added{ed} an average value for the cosine, i.e., $\cos(i) = 0.5$.

$F_{disk}$ is fitted simultaneously with $F_{WD}$. We used the estimated values of $R_{WD}$, $M_{WD}$,  and $\dot M$ from the hard X-rays fitting. 
$T_{WD}$, $T_{in}$, and $T_{out}$ are free parameters in the fit of infrared and optical SED.

\section{An accreting WD model for 4U~0142+61} 
\label{sec:fit}

In this \replaced{S}{s}ection, we describe the fit procedure of the dereddened and deabsorbed \fu SED~(Section \ref{sec:data}). We assume\added{d} the model presented in Section~\ref{sec:model}, which in based in \replaced{an}{a magnetic} accreting WD model for 4U~0142+61. We also discuss the consequences of the derived parameters for \replaced{an}{a} WD scenario. 

\subsection{Fitting 4U~0142+61 SED}
\label{sec:ajuste}

As the model parameters for each spectral region are not the same, we opted to fit spectral regions separately. Doing this, we could also constrain some parameters in a simpler way than doing an overall fitting. First, we fit\added{ted} the hard X-rays independently of other parts of the SED. After that, we fitted the soft X-rays considering the contribution of hard X-rays bremsstrahlung in this energy range. Finally, we fit\added{ted} the optical/IR emission taking all the previously fitted components and parameters into account. 
\added{The distance to \fu is considered equal to 3.78~kpc (see Section \ref{sec:distance} in all calculations.} 

We \deleted{have} used \added{the} Markov Chain Monte Carlo \added{method} \cite[MCMC -][]{2010CAMCS...5...65G} to estimate the parameters and their uncertainties. \replaced{We adopt as the fit}{The adopted} figure of merit \added{is} the ratio between $\chi^2$ and the degrees of freedom (dof). The parameters of the fit and the resulting SED are shown in Table~\ref{tab:results} and Figure\deleted{s}~\ref{fig:SED}.  

The fit quality of the hard X-rays increases with \added{the} bremsstrahlung temperature\deleted{s}, but \replaced{we have}{there is} a limit imposed by the maximum mass a WD can have since the temperature is related to the WD mass by
equations \ref{tempnovo} and \ref{eq:temp_brem}. The highest temperature we can reach for the limiting mass of $1.41$~$M_{\odot}$ and radius of 1,021~km \citep{2018GReGr..50...38C} is 674.5~keV, which results in a $\chi^2/dof = 0.84$ for the hard X-rays emission. 
If we use a smaller mass, for instance,  $1.36$~$M_{\odot}$, 
we obtain $\sim 250$~keV, which gives $\chi^2/dof=1.04$.

\replaced{Thus, we fixed $M_{WD}=1.41 M_{\odot}$  and $R_{WD} = 1,021~km$, which implies a temperature of 674.5~keV, in order to guarantee the best fit of \fu hard X-rays spectrum for a WD scenario. We fit the hard X-rays data using equation~\ref{eq:brem_emission}. This equation fits the hard SED using solely $N$ (see equation~\ref{eq:emission}), which is a free parameter, $T_{brem}$ (fixed at 674.5~keV) and distance $d$ (fixed at 3.78~kpc). Thus, $N$ is the only free parameter for the MCMC fitting of the hard X-rays fit.}
{We fit the hard X-rays data using equation~\ref{eq:brem_emission}. The only free parameter is $N$ (see equation~\ref{eq:emission}), since $T_{brem}$ is fixed at 674.5~keV (see the above paragraph).}

After modelling the hard X-rays, we \replaced{find the best}{performed the} fit \replaced{for}{of the} soft X-rays. The bremsstrahlung component \replaced{is}{was} also included in the fit\deleted{ the soft X-ray SED}. To increase the quality of the fit, we use\added{d} two black bodies components, which can have different temperatures and radii. 
The flux for each hot spot is given by equation~\ref{eq:flux_spot}\replaced{. In this case,}{, in which} $T_{spot 1}$, $R_{spot 1}$, $T_{spot 2}$, and $R_{spot 2}$ are free parameters and $d$ is fixed. In Table~\ref{tab:results}, we present the parameters for hard and soft X-rays fit and Figure~\ref{fig:SED} shows the fitted model. 

\added{Two spots with different temperatures are necessary to represent the multi-temperature characteristic of the soft X-ray emission: if we use only one blackbody, the soft X-rays are not well fit.  The two-spot approximation can be a mathematical simplification for a single spot with a temperature gradient, which is consistent with the smooth soft X-ray emission - in opposition to a double-peaked soft X-ray emission. On the other hand, the two-spot model can reflect the presence of two accretion columns, which would imply two bremsstrahlung components in hard X-rays. However, due to the similar power-law behavior of the bremsstrahlung emission for \mbox{T $<$ 100~keV}, two components with slightly different temperatures are indistinguishable. Therefore, we could not differentiate between one or two post-shock regions using the observed SED in hard X-rays.}

Using the parameters obtained in X-rays fits, we can derive \deleted{the} some important quantities of the system\deleted{, which is done below}. First, we estimate\added{d} $L_{brem}$, which is the accretion column luminosity\replaced{ due to accretion estimated}{,} by integrating $F_{brem}$ in all energies.
Using the estimated $L_{brem}$, we calculate\added{d} $\dot M$ (see equation~\ref{eq:mdot}). We consider\added{ed} the area of the column equal to the soft X-rays emitting area, which gives an average radius for $R_{brem}$ of 14.02~km. Thus, we were able to estimate $n_e$ using equations~\ref{eq:emission} and \ref{eq:ne}. Knowing $N$, $n_e$, and $R_{brem}$, we have the value of $H_{brem}$\replaced{, which is 48\% of the upper limit given by equation~\ref{eq:height}.}{The downward bremsstrahlung photons illuminates not only an area equal to the footprint of the accretion column, but also a surrounding circular ring. In other words, we should have $R_{brem}$ of the same order, but still smaller, than $\sqrt{R_{spot1}^2+R_{spot2}^2}$.  Nonetheless, we can have higher $H_{brem}$ and, consequently, smaller $R_{brem}$, since $H_{brem}$ is only 48\% of the upper limit given by equation~\ref{eq:height}}.

To fit the optical and IR emission, we use\added{d} the WD photosphere blackbody and the debris disk\replaced{. The fluxes of the WD photosphere and of the disk}{, whose fluxes} are given by equations~\ref{eq:flux_photosphere} and \ref{eq:flux_disk}, respectively. We use\added{d the} same values of $R_{WD}$ derived from the bremsstrahlung fit. Moreover, we included the tail of the bremsstrahlung component in the fit. 
We \replaced{do}{did} not consider the presence of hot spots in the fit of the optical/IR \added{data} because its emission \deleted{in this energy range} is negligible \added{in this energy range}. The best fit for the optical and IR is presented in Table~\ref{tab:results} and in Figure~\ref{fig:SED}. \cite{2000Natur.408..689H,2004A&A...416.1037H} data
follow a different slope compared with mid and near-IR data
(\citealt{2006ApJ...652..576D}; \citealt{2006Natur.440..772W}) whereas the recent optical data from \cite{2016MNRAS.458L.114M} fits more smoothly with the IR emission. Due to those differences, we perform the optical fit considering only the data from \cite{2016MNRAS.458L.114M} but present both data in Figure~\ref{fig:SED}.b, in order to show that the fit is consistent with both optical ensembles.

Figure~\ref{fig:SED} also shows the upper limits of gamma-ray emission from \fu obtained using the CGRO COMPTEL instrument \citep{denhartog2008}. Hence, the proposed
model is able to fit consistently all observational data of \fuu. 

\begin{table*}[tbhp]
\begin{center}
\caption{Parameters of the fitting of \fu in the accreting WD model.}
\label{tab:results}
\begin{tabular}{ c  c  c }
\hline
\hline
Parameter & Description& Value \\
\hline
\multicolumn{3}{c}{{\bf X-rays}} \\
\hline
\multicolumn{3}{c}{{\sc Fixed parameters}}  \\[0.5ex]

$d$ & distance of \fu  & 3.78~kpc\\
$N_H$& columnar density of hydrogen & 6.4~$10^{21}$~cm$^{-2}$ \\
$T_{brem}$& temperature of the emission for the accretion column & 674.5~keV \\
$M_{WD}$ &  WD's mass &1.41~$M_{\odot}$ \\
$R_{WD}$ &  WD's radius &1,021~$10^5$~cm \\
\hline
\multicolumn{3}{c}{{\sc Fitted parameters}}  \\[0.5ex]
$N$ & normalization parameter & 3.30 $\pm$ 1.09~$10^{56}$~cm$^{-3}$ \\
$\chi_{brem}^2/dof$ & reduced chi square for the hard X-rays &0.85 \\
$T_{spot1}$ & temperature of the spot 1  &0.632 $\pm$ 0.033 keV \\
$R_{spot1}$& radius of the spot 1 &2.35$\pm$0.45~$10^5$~cm  \\
$T_{spot2}$ & temperature of the spot 2 &0.337$\pm$0.012~keV\\
$R_{spot2}$ & radius of the spot 2 &13.83$\pm$0.73~$10^5$~cm\\
$\chi^2/dof$ & reduced chi square for the soft X-rays &1.06\\
\hline
\multicolumn{3}{c}{{\sc Derived parameters}}  \\[0.5ex]
$T_{shock}$& temperature of the shock front & 1,053.9~keV \\
$L_{brem}$ & bolometric luminosity due to bremsstrahlung & 6.29~$10^{35}$~erg~s$^{-1}$ \\
$\dot M$ & accretion rate & 3.43~$10^{17}$g~s$^{-1}$ \\
$R_{brem}$& radius of the hard X-ray emission &14.03~$10^5$~cm \\
$H_{brem}$ & height of the accretion column & 1.27~$10^5$~cm \\
$n_e$ & electrons number density & 2.05~$10^{19}$~cm$^{-3}$ \\
\hline
\multicolumn{3}{c}{{\bf Optical/IR}} \\
\hline
\multicolumn{3}{c}{{\sc Fitted parameters}}  \\[0.5ex]
$T_{WD}$ & WD's effective temperature  & 9.4$\pm$7.3~$10^4$~K \\
$T_{in}$ & inner temperature of the debris disk & 1,991$\pm$16~K\\
$T_{out}$ & outer temperature of the debris disk & $285\pm200$~K \\
$A_d$ & albedo of the disk & $0.985\pm0.003$ \\
$\chi^2/dof^{a}$ &reduced chi square for the optical/IR range & 2.45\\
\hline
\multicolumn{3}{c}{{\sc Derived parameters}} \\[0.5ex]
$R_{in}$ & inner radius of the debris disk &   $2.35\pm0.03$~$R_{\odot}$\\
$R_{out}$ & outer radius of the debris disk & $31^{+127}_{-16}$~$R_{\odot}$ \\
\hline
\end{tabular}
\end{center}

\textbf{Notes.}The fixed parameters were derived before the fit by independent methods. For the IR/optical fit all the X-rays parameters are considered fixed, therefore, $R_{WD}$ is not a fitted parameter for this range of energy. The $1\sigma$ uncertainties for the last digit for the fitted parameters are between parenthesis.
\explain{Due to new bremsstrahlung emission expression, the changes below were done. They are not annotated in the table itself, because the table runs off the page in this case. 
(1) The WD's effective temperature was 9.5$\pm$7.4~$10^4$~K. 
(2) The inner temperature of the debris disk was 1,947$\pm$17~K.
(3) The outer temperature of the debris disk was $295\pm205$~K.
(4) The inner radius of the debris disk was   $2.43\pm0.03$~$R_{\odot}$.
(5) The outer radius of the debris disk was $30^{+114}_{-16}$~$R_{\odot}$.
}

a: The estimate of $\chi^2/dof$ in the optical/infrared range did not take into account the {\it $K_s$}-band, since it presents a distinct trend and would disturb the fit of other bands.
\end{table*}

\subsection{Discussion on the resulting parameters of \fu}

\subsubsection{Post-shock region} 
\label{sec:discussion_shock}

\cite{2014RAA....14..673W} tried for the first time to fit the hard X-rays of \fu with a bremsstrahlung component and found a poor fit. However, they used \deleted{the} XSPEC 12.6.0q, which uses a non-relativistic Gaunt factor derived from \cite{1961ApJS....6..167K} and \cite{1975ApJ...199..299K}. Conversely, we used the relativistic Gaunt factor from \cite{1998ApJ...507..530N}, which, depending on the energy range, can differ from the non-relativistic prescription by orders of magnitude. Moreover, we use a composition of carbon opposed to the 92.5\% of hydrogen and 8.5\% of helium and applied the relativistic correction for high temperatures. All those differences in the methodology allow us to fit the data with a bremsstrahlung component that has a natural cutoff \added{at high energies}, in agreement with the observed \replaced{cutoff}{data}~(see Fig.~\ref{fig:SED}).
 
The absence of $H\alpha$ emission in the optical spectrum of \fu was interpreted by \cite{2016MNRAS.458L.114M} as an evidence against accretion. However, \replaced{this}{the} carbon composition of the accreted mass  naturally explains the absence of $H\alpha$  emission.

To fit the data, we need a \deleted{high} value \replaced{of}{for the} temperature \replaced{for}{of} the accretion structure\deleted{,} around 670~keV. Such a high temperature is not observed for any known cataclysmic variable. However, high values are theoretically possible for a massive WD (see equation~\ref{tempnovo}). In addition, an in-falling matter composed by carbon and oxygen increases the mean molecular weight, which also increases the temperature. Also, D. Belloni~(private communication) has implemented a shock solution for accreting WDs. Their results show that shock temperature\added{s} as high as 1000~keV can be found for massive WDs accreting C-O material, which corroborates the temperature of our fitting

\subsubsection{WD temperature and age}
\label{sec:wd}

The temperature of the WD photosphere for our
model for 4U~0142+61 is high (\replaced{$T_{WD}$~=~$9.5 \times10^4$~K}{$T_{WD}$~=~$9.4 \times10^4$~K}).
\cite{2015A&A...584A..19W} present two extremely hot WDs, H~1504+65 ($T_{WD}$~=~$2.0\times10^5$~K and $M$~=~0.83~$M_{\odot}$) and RX~J0439.8-6709 ($T_{WD}$~=~$2.5\times10^5$~K and $M$~=~0.86~$M_{\odot}$). Both of them are hotter than our models. Hence, our findings are consistent with temperatures that we do see in WDs.

From the effective temperature, we can estimate the WD age. The Mestel cooling law \citep{1952MNRAS.112..583M} was the first attempt of such estimate. \cite{2003ApJ...589..179H} present an improved version of \replaced{his}{that} cooling law, which for \replaced{t}{ages}~$<$~9,000~Myr is

\begin{equation}\label{eq:cool}
    L = \frac{300MZ^{0.4}}{\big[A(t+0.1)\big]^{1.18}},
\end{equation}

\noindent where $L$ is the WD luminosity in solar units, $M$ is the WD mass in solar units, $A$ denotes the average atomic number, $t$ is the age in Myr, and $Z$ is the metallicity. We consider a core composition of 60\% carbon and 40\% oxygen, and $Z$~=~0.001 \citep{2013ApJ...772L..24R}, which results in a cooling age of 5.5~Myr\deleted{, considering the errors in $T_{WD}$ estimates}.

\subsubsection{Debris Disk}\label{sec:disk}

The inner temperature of the debris disk is 1\,947~K \replaced{1\,947~K}{1\,991~K}
(see Table~\ref{tab:results}), larger than the silicate sublimation temperature ($T_s$) of about 1300~--~1500~K \citep{2003ApJ...591.1220L}. However, this $T_s$ is based on the solar abundance and is mainly used to model protoplanetary disk of young stars. \cite{2012ApJ...760..123R} argue that those values of $T_s$ \replaced{provide underestimated values}{are underestimates} of $T_{in}$ for disk around WDs\added{,} once the composition and evolution of these disks are distinct from those around young stars. In fact, some WDs have $T_{in}$ larger than 1500~K, such as HE~1349$-$2305~\citep{2012ApJ...749..154G} and GD~56~\citep{2009ApJ...694..805F}, both with $T_{in}=1700$~K. Moreover, according to \cite{2012ApJ...760..123R}, $T_{in}$ is larger for WD with higher accretion rates and $T_{WD}$, which \replaced{justify}{agrees with} the larger $T_{in}$ of \fu compared to  $T_{in}$ of other isolated WDs.

Once we consider the debris disk opaque, we can estimate the minimum mass of the dusty disk using the approach from \cite{2003ApJ...584L..91J}. The  surface area of the dusty region can vary from $A\sim 10^{25}$ to $A\sim 10^{27}$~cm$^{2}$.
\cite{2012ApJ...760..123R} state that the particle sizes in debris disk around WD vary from 0.03 to 30~cm. 
Thus, if we consider 10~cm-diameter particles, the opacity is $0.05$~g$^{-1}$\,cm$^{2}$ at 15$~\mu$m~\citep{1992A&A...261..567O}, resulting in a minimum dust mass around $10^{26}$~--~$10^{28}$~g.

Once the disk is formed by gas and dust, we should also estimate the mass in gaseous form. \citet{2016ApJ...828...46A} suggest that protoplanetary disks can have a gas-to-dust mass ratio from $10^0$ up to $10^3$. Assuming the maximum value, the total mass of the 
disk would be around $10^{29}$~--~$10^{31}$~g.

\replaced{However, this}{The above} value does not consider the mass in the internal gas disk. In view of all approximations and uncertainties, this value should be regarded as a crude estimate of the minimum disk mass. As we will discuss in Sec.~\ref{sec:origin_evol}, the expected total mass of a disk resulting from the merger of two CO WDs is about~$10^{32}$~g.  

\subsubsection{Gaseous disk}
\label{sec:depth}

Our fit is based on the assumption that the gaseous disk is optically thin and does not contribute to the optical and infrared emission of the system. In this section, we demonstrate that it is a valid assumption.

\replaced{For this we use}{We adopted} the thin disks model proposed by \cite{2002apa..book.....F}, with some adaptations to fit our needs. Below we present the main equations, which are not dependent of any assumption on the optical depth. Considering a geometrically thin steady disk, the optical depth $\tau$ is:

\begin{equation}
    \tau (R) = \Sigma(R) \kappa_R(R),
\end{equation}

\noindent where $\kappa_R$ is the Rosseland mean opacity and \replaced{$\Sigma$}{$\Sigma$(R)} is the disk surface density. Assuming that \replaced{the viscosity follows an alpha-description as proposed by Shakurav \& Sunyaev}{the viscosity parameter $\alpha$ follows the \cite{1973A&A....24..337S} prescription} and that the disk is geometrically thin, we can write \replaced{$\Sigma$}{$\Sigma$(R)} as \citep{2002apa..book.....F}:

\begin{equation}
   \Sigma (R)  = \frac{\dot M}{3\pi c_s\alpha H(R)}\left[1-\left(\frac{R_{WD}}{R}\right)^{0.5}\right],
\end{equation}

\noindent where $c_s$ is the sound speed, $H(R)$ is the height of the disk in \replaced{a}{the} radius $R$\deleted{, and the viscosity parameter $\alpha$ follows the \cite{1973A&A....24..337S} prescription}. We must point out that there are some discussions about the validity of the $\alpha$-parametrization for the outcome disk of a merger of WDs~\citep{2018ApJ...857..134B}. 
\replaced{It}{Its use} is justified by the lack of other applicable theories and the widespread use of such prescription in disks surrounding isolated WDs and pre main-sequences stars\deleted{ by other authors}.

Moreover, we can estimate $c_s$ from the height \replaced{$H$}{$H$(R)} and $R$ using~\citep{2002apa..book.....F}:

\begin{equation}
c_s (R) =H(R)\frac{\left(GM_{WD}\right)^{0.5}}{R^{3/2}}.
\end{equation}

Following the methodology from \cite{2002apa..book.....F}, we use the Kramer's law to estimate \replaced{$\kappa _R$}{$\kappa _R$(R)}. However, we consider\added{ed} a carbon~(or entirely metallic) composition instead of the solar one, which gives~\citep{hansen2004stellar}:

\begin{equation}
    \kappa_R (R) =4\times 10^{25} \rho(R) T_c(R)^{-7/2},
\end{equation}

\noindent where \replaced{$T_c$}{$T_c$(R)} is the central temperature and \replaced{$\rho$}{$\rho$(R)} is the volumetric density, given by~\citep{2002apa..book.....F}:

\begin{equation}
    \rho (R) = \frac{\Sigma(R)}{H(R)}.
\end{equation}

In contrast to \cite{2002apa..book.....F}, we calculated \replaced{$T_c$}{$T_c$(R)} considering that the radiation pressure is negligible, i.e., the pressure has only the gas component. Doing this, we obtain:

\begin{equation}
    T_c (R) =\frac{c_s(R)^2 \mu m_p}{k},
\end{equation}

\noindent where $m_p$ is the proton mass.

It is not possible to estimate the optical depth solely by the equations above, since there is more parameters than equations. The missing equation \replaced{from}{in} the model of \cite{2002apa..book.....F} correlates \replaced{$T_c$}{$T_c$)R)} with \replaced{$\tau$}{$\tau$(R)}. However, this equation is \deleted{derived} based on the assumption of \replaced{a}{an} optically thick disk. Since\deleted{,} we \deleted{are trying to find}{aim to verify} whether the disk is optically thin or not\added{,} we cannot consider an equation for an optically thick disk. Thus, we propose a new equation for the system to be solvable.

We assume\added{d} $H(R)/R\sim0.1$, following the simulations of \cite{2012ApJ...746...62R} for a disk around a WD produced in a merger. This hypothesis is corroborated by the fitting of the SED of FU Ori, which also resulted in $H/R\sim 0.1$ \citep{1991ApJ...383..664K}. This result was later used by \cite{1993ApJ...407..219H} to estimate the optical depth of Ae Herbig stars disks, which could be optically thin for accretion rates of $\sim10^{-9}$~M$_{\odot}$yr$^{-1}$. Thus, supported by results related to merger simulations and to pre-main-sequence disks, we fixed the relation between $R$ and $H$ by \deleted{the} equation~\ref{eq_H}. Hence, we can calculate the optical depth of the gaseous disk.

\begin{equation}
    H(R)=0.1R
    \label{eq_H}
\end{equation}

We arbitrate\added{d} that the disk is optically thick if the optical depth \replaced{$\tau>1$}{is larger than 1}. Using the above equations, we obtain\added{ed} $\tau\leq1$ for $\alpha\geq 0.15$, which is consistent with $\alpha=$~0.1~--~0.4 for accretion disks~\citep{2007MNRAS.376.1740K}. \added{Hence, the gaseous disk of our model is optically thin.}

\subsubsection{Accretion rate}

Our model predicts an accretion rate of $\dot M \sim 10^{17}$g~s$^{-1}$, which is much higher than $\dot M\sim 10^{5}$~--~$10^{11}$g~s$^{-1}$ for isolated WDs~\citep{2014A&A...566A..34K,2018MNRAS.474..947F}. Conversely, our $\dot M$ is similar to pre~main-sequence stars, such as T Tauri\deleted{s} and Ae/Be Herbig\replaced{s}{ objects}~\citep{2008ApJ...676L.139V,2011A&A...535A..99M}. This is consistent with our hypothesis that the \fu disk is similar to protoplanetary disks (see Sec. \ref{sec:multi_temp_disk}). 

Analyzing the population of T Tauri\replaced{s}{ objects}, \citet{2008ApJ...676L.139V} proposed a correlation between the disk mass and the accretion rate of those objects given by $\dot M_{disk}=10^{-7}{M_{disk}}^{1.1}$. Hence, expanding this behavior for isolated WDs, the high difference between the accretion rate of \fu and isolated WDs can be qualitatively explained by the huge difference between their disk masses~(see Tab~\ref{tab:disk_mass}).

Table~\ref{tab:disk_mass} compares the accretion rates for \mbox{WD~G29-38} \added{(a isolated WD)} and \fu obtained from X-ray emission, $\dot M_{XR}$, with the estimated accretion rate calculated using $M_{disk}$. Interestingly, both values are very close, giving some plausibility for the proposed relation.

\begin{table}[hbtp]
    \begin{center}
        \caption{Comparison between $\dot M_{XR}$ for \fu and \mbox{WD~G29-38}
        and the estimated value based on the relation between the disk mass and the accretion rate, which is denoted by $\dot M_{disk}$. We consider the same dust-to-gas ratio from Sec.~\ref{sec:disk}.
        }
    \label{tab:disk_mass}
    \begin{tabular}{c|c c c c}
    \hline
    Object & $M_{dust}$ &$M_{disk}$ &  $\dot M_{disk}$ & $\dot M_{XR}$  \\ \hline
       \fu  & $\sim10^{29}$\tablenotemark{\scriptsize{a}}&$\sim10^{32}$&$2\times 10^{17}$ & $3.4\times 10^{17}$\tablenotemark{\scriptsize{a}} \\
        WD G29-38 & $\sim10^{23}$\tablenotemark{\scriptsize{b}}&$\sim10^{26}$&$6\times10^{10}$  &$7 \times 10^{10}$\tablenotemark{\scriptsize{c}}  \\ \hline
    \end{tabular}
    \end{center}
    
    \tablenotetext{a}{Our work}
     \tablenotetext{b}{\citet{2003ApJ...584L..91J}}
    \tablenotetext{c}{\citet{2018MNRAS.474..947F}.}
\end{table}

\subsection{Optical pulsed fraction}

The main goal of this paper is to propose a plausible model to the quiescent spectrum of \fu in a WD scenario. However, our model is also consistent with the \fu flux modulation, as we show below.

\fu has periodic variability in the X-rays and optical wavelengths~\citep{denhartog2008,2002Natur.417..527K}. The X-ray pulsed emission could be explained by the change of the observer view of the the magnetic accretion structure and of \added{the} hot spots on the WD surface along the WD rotation. \added{Regarding the optical pulsation, the accreting magnetic WD model has the cyclotron emission as a common explanation. However, we did not include this radiative process in our model.} The non-negligible \deleted{ contribution of the} bremsstrahlung \added{contribution,} \deleted{from the post-shock region} \added{which is still optically thin} in optical wavelengths (see Fig.~\ref{fig:SED})\added{, can} also explain qualitatively the flux moduation in \replaced{this spectral region}{the optical regime}. 

Below we show that the optical pulsed fraction, PF, is also quantitatively consistent with the proposed scenario. We adopt the PF definition of \citet{2002Natur.417..527K}\deleted{, which is}

\begin{equation}\label{eq:pulsed_fraction_cvr}
    PF = \frac{F_{max}-F_{min}}{F_{max}+F_{min}},
\end{equation}

\noindent in which $F_{max}$ and $F_{min}$ are, respectively, the maximum and minimum fluxes during a rotation. 

\replaced{The}{In our model, the} \deleted{case of} maximum possible optical PF \deleted{in our model} occurs if the post-shock region is completed self-eclipsed by the WD. In this case, we can write

$$F_{min} = F_{WD}+F_{disk},$$

\noindent and

$$F_{max} = F_{WD}+F_{disk}+F_{brem}.$$

Thus, the maximum PF in \deleted{the} \textit{I}-band is \replaced{20\%}{28\%},
which is very close to the observational PF of $27^{+8}_{-6}$\% \deleted{for this band} \citep{2002Natur.417..527K}.

\subsection{Comparison with previous models}

For the magnetar model, the X-rays SED of \fu was successfully modelled by \cite{2014ApJ...786L...1H} using the phenomenology proposed by \cite{2013ApJ...762...13B}.
In this scenario, the IR emission is attributed to a passive irradiated disk~\citep{2006Natur.440..772W}. The optical \replaced{are}{emission is} supposed to be of magnetospheric origin~\citep{2006Natur.440..772W}, however, it lacks a quantitatively explanation for such emission. 

Similarly to the magnetar model, the accreting NS model can reproduce the X-rays emission \citep{2015MNRAS.454.3366Z}. But, for this model, there are some attempts to fit the optical and IR data using a fallback disk.  \citet{2000Natur.408..689H} and \citet{2000ApJ...541..344P} were unable to fit \fu optical data using a fallback disk with $T~\alpha$~$R^{-3/7}$. Using a model in which $T~\alpha$~$R^{-1/2}$ and $F~\alpha$~$\nu^{-1}$, \citet{2006ApJ...649L..87E} and \citet{2007ApJ...657..441E} were able to fit the optical/IR data.

The X-rays of \fu in the WD pulsar model is due to a pulsar-like emission. \added{ However, we are not aware of any attempt to fit \fu X-rays SED in this model.} Recently, \citet{2017MNRAS.465.4434C} have inferred \deleted{the values of} the structure parameters, magnetic field, rotation period, and spin-down rates of a WD pulsar death-line\replaced{, showing}{. They have shown} that WDs above the death-line emit blackbody radiation in the soft X-ray band via the magnetic polar cap heating by back flowing pair-created particle bombardment. Moreover, the WD pulsar model has been used to explain the emission of other objects, such as AR~Scorpii~\citep{2016ApJ...831L..10G}. \deleted{However, we are not aware of any attempt to fit \fu X-rays SED using the WD pulsar prescription. }
 
The optical/IR modelling of \fu in the WD pulsar model uses the same components of our model: a debris disk modelled by equation \ref{eq:flux_disk} and \replaced{the}{a} photosphere emitting \deleted{such} as a blackbody~\citep{2013ApJ...772L..24R}. For this model, $R_{WD}=4.2\times10^8$~cm, $M_{WD}=1.1$~$M_{\odot}$, and $T_{WD}=1.31\times10^5$~K. In comparison, our estimates are more extreme\replaced{, with a}{. We obtained} smaller $R_{WD}$\replaced{ and}{,} higher $M_{WD}$\added{,} and roughly similar $T_{WD}$. Apart from the different data set, we used the derived $R_{WD}$ from the hard X-rays in order to build a model that is consistent in all wavelengths, whereas \cite{2013ApJ...772L..24R} use the optical/IR data alone to derive the WD parameters, which results in more standard parameters. 
Thus, even though the fits have distinct $R_{WD}$ and $T_{WD}$, \fu is thought to be a young WD in both cases. 

The nature of the optical pulsed emission is challenging for all previous models. For a NS scenario, \cite{2002Natur.417..527K} pointed out that the observed fraction is higher than the expected by a disk \replaced{irradiation}{irradiated by} X-rays. Thus, \citet{2002Natur.417..527K} proposed that such pulsation could be magnetospheric in the magnetar model. However, no component of the \cite{2014ApJ...786L...1H}'s model extends up to the optical range. The NS accreting model does not present estimates \replaced{in}{for} the expected optical pulsed fraction, even though \cite{2004ApJ...605..840E} state that the pulsed fraction could be cause\added{d} by outer gaps in the pulsar magnetosphere operating with \replaced{a disk inside}{an internal disk}. To our knowledge, there is no attempt to explain the optical pulsation in the context of the WD pulsar model.

\section{Spin-down, propeller regime, and the magnetic field} \label{sec:spin-down}

4U~0142+61 is slowing down. Thus, in this section, we inferred the magnetic field of 4U~0142+61 to reproduce the spin-down in an accreting regime.

The corotational radius\added{, $R_{c}$ (equation~\ref{eq:corotational}),} is the disk position in which the particles rotational velocity is equal to stellar rotation. According to \cite{2005ApJ...623L..41E}, $R_c$ \deleted{(equation~\ref{eq:corotational})} must be larger than both $R_{WD}$ and $R_m$ for the system to be in the accreting regime. Moreover, $R_m$ must be \replaced{higher}{larger} than $R_{WD}$ for the accretion to be magnetic and create spots. 

If we consider only the effect of matter falling onto the WD, we \replaced{would}{will} have \replaced{spin-up}{spin up}. However, \replaced{it is possible for the WD to spun-down}{the WD can spin down} in the accretor regime if other spin-down mechanisms compensate \replaced{that}{the} increase due to accretion, such as the coupling between the magnetic field and the disk or the misalignment \replaced{of the magnetic field with the rotation axis}{between the magnetic field and rotation axes} \citep{1987A&A...183..257W,2012ApJ...749...25G}.

\cite{2012ApJ...749...25G} argue that the alignment of the \added{dipole} magnetic field with the spin axis can be the main cause of the spin-down for merger remnants. That spin-down effect is more prominent on WDs that are the product of a merger of two WDs with different masses, once the misalignment between the magnetic field and the spin \replaced{axis}{axes} is \replaced{bigger}{larger} in that case \citep[see][for details]{2012ApJ...749...25G}. This spin-down can be estimated by:
 
\begin{equation}\label{eq:spin-down_pulsar}
    \dot \Omega_{mag} =- \frac{2 \Omega^3 \mu^2}{3Ic^3}\sin^2\beta,
\end{equation}

\noindent where $\mu$ (G~cm$^{3}$) is the magnetic moment, $I$ (g~cm$^{2}$) is the moment of inertia, and $\beta$  represents the angle between the magnetic field \replaced{lines and the axis of rotation}{and rotation axes}. 

The accretion and coupling spin change can be calculated by~\citep{1987A&A...183..257W}:

\begin{equation}\label{eq:spin-down_accretion}
   \dot \Omega_{acc} = \frac{2\pi\dot M R_c^2}{PI}n(\omega_s),
\end{equation}

\noindent where $n(\omega_s)$ can be obtained in \cite{1987A&A...183..257W}, for $\omega_s=R_m/R_c$. In this case, the equation account\added{s} for both the components\replaced{,}{:} the coupling between the magnetic field and the disk and the \replaced{matter falling into to}{accretion onto} the central remnant. In this model, \replaced{it is possible to spun-down}{the WD can spin down} if the \replaced{relation}{ratio} between the magnetosphere radius and the corotational radius is in the range of 0.971 to 1.0 \citep{1987A&A...183..257W}. \replaced{For $R_m$, we adopt the estimate from}{To estimate $R_m$, we adopted} \citep{1989ApJ...341..327F},

\begin{eqnarray}
\label{eq:alfven}
\nonumber \frac{R_{m}}{R_{WD}}\simeq13.4\bigg[\frac{B(1+3\sin^2\beta)^{1/2}}{3 \times 10^7} \bigg]^{4/7}\bigg(\frac{f}{10^{-3}}\bigg)^{2/7}\times \\
\times\bigg(\frac{M_{WD}}{M_{\odot}}\bigg)^{-8/21}\bigg(\frac{\dot  M}{10^{16}\text{~g~s}^{-1}}\bigg)^{-2/7},
\end{eqnarray}

\noindent where $B$ is the dipolar magnetic field. 

The total spin-down is given by  

\begin{equation}\label{eq:spin-down}
    \dot P = \frac{-P^2}{2\pi}\big(\dot \Omega_{acc}+\dot \Omega_{mag}\big).
\end{equation}

\fu must have $R_c$~$>$~$R_m$~$>$~$R_{WD}$ and a spin-down \replaced{in the order of}{around} $10^{-12}$~s\,s$^{-1}$ for our model to be \replaced{viable}{feasible}. Moreover, the inferred value of magnetic field to reproduce the spin-down must be consistent with a bremsstrahlung scenario. Depending on the nature of the WD and its accretion structure, different types of \replaced{radiative processes}{cooling} could be predominant \citep{1979ApJ...234L.117L,1982ApJS...48..239K}. For bremsstrahlung \replaced{dominate instead of}{dominates over} cyclotron emission, the magnetic field has to be smaller than $6 \times 10^6 \,(L_f/10^{36} $erg\,s$^{-1})^{2/5}$, where $L_f$ is $L_{brem}$ divided by $f$. Considering $R_{brem}$ and $R_{WD}$, we have $f \sim 1.88\times 10^{-4}$, which gives an upper limit for $B$ of $2.03\times10^8$~G. 

Adopting a moment of inertia \replaced{in the order of}{equals to} $10^{48}$~g~cm$^{2}$ and considering that the pulsar magnetic dipole moment is misaligned with \replaced{its}{the} spin axis by an angle $\beta=90^{\circ}$, we \replaced{have an}{obtain} $\dot P_{mag}$ of $5.3\times10^{-15}$~s~s$^{-1}$\added{, which is=} much smaller than the observed spin-down of \fu. Therefore, \replaced{$P_{mag}$}{the contribution from the misalignment of the magnetic field and rotation axes} is not able to reproduce the spin-down and can be neglected.
Thus, we need to reproduce the spin-down solely by \added{the coupling disk-magnetic filed, see} equation~\ref{eq:spin-down_accretion}, which \replaced{gives}{imposes} $0.971<R_m/R_c<1$. \replaced{This way, if we reproduce the spin-down by equation~\ref{eq:spin-down_accretion},}{Therefore, if this proof is successful,} we will also prove the assumption that $R_c\sim R_m$\replaced{. It's possible to}{, which was used in the } estimate $T_{shock}$ by equation~\ref{tempnovo}. For $\beta=90^{\circ}$\added{,} we must have $B=2.82\times10^7$~G to reach \replaced{the}{$\dot P =$} $2\times10^{-12}$~s\,s$^{-1}$. Conversely, if we consider $\beta=0^{\circ}$, we must have $B=5.63\times10^7$~G. Thus, \replaced{the magnetic field range in the WD accreting model is}{we obtain} $2.82\times10^7<B<5.63\times10^7$~G. Those values are consistent with the upper limit imposed by the bremsstrahlung emission. Moreover, this spin-down requires $R_m/R_c=0.994$, which is extremely close but still consistent with the \replaced{criteria}{condition} to be in the accreting regime.

\replaced{Moreover, we compare and contrast these magnetic field inferred with respect to the one observed from }{We can compare the above values of magnetic fields with those observed in} magnetic WDs. Polars magnetic field varies from 7~MG~(V2301~Oph) up to 240~MG~(AR~UMa)\added{,} whereas \replaced{Intermediate Polars}{intermediate polars} have magnetic field\added{s} of $\sim$~4 -- 30~MG, with  the highest value of $\sim$~32~MG for V405~Aur \citep{2015SSRv..191..111F}. \replaced{For i}{I}solated WDs\replaced{, the}{have} magnetic field\added{s} \deleted{is} in the range $10^3$~G \added{--} $\sim 10^9$~G. \replaced{Close to $10^7$~G, we have several examples of isolated WDs such as WD~0806+376, with a field of $3.97\times10^7$~G and WD~1017-367, with $B=6.5\times10^7$~G}{There are several examples of isolated WDs that have B around $10^7$~G. We cite some: WD~0806+376, with a field of $3.97\times10^7$~G, and WD~1017-367, with $B=6.5\times10^7$~G} \cite[see][for \replaced{another}{other} examples of magnetic WDs]{2015SSRv..191..111F}. Thus, our estimate of magnetic field is consistent with \replaced{all classes of}{the observed values in} magnetic CVs.

\section{Possible origin and evolution of the object}
\label{sec:origin_evol}

In this accreting WD scenario, 4U~0142+61 is a fast-spinning, isolated, magnetic, hot, and extremely massive WD. Even though those characteristics are very uncommon for \replaced{an}{a} WD, sources with similar characteristics have already been observed. RE~J0317-853, for example, is in an binary system without any interaction with the secondary. This object has a period of 725.4~s, \replaced{a}{an} estimated mass of 1.35~$M_{\odot}$, an effective temperature of $\sim$ 50.000~K, and a magnetic field of $\sim340$~MG \citep{1995MNRAS.277..971B}. The most plausible origin for that source is the merger of two less massive CO WDs \citep{1997MNRAS.292..205F}, which is the same origin proposed to \fu in the WD pulsar model~\citep{2013ApJ...772L..24R}. 

There are some arguments in favor of a merger origin for massive and magnetic WDs. For instance, merger is one of the most plausible scenarios to explain why magnetic isolated WDs are usually more massive than the non-magnetic ones \citep{2012ApJ...749...25G}. In fact, all \added{the} 4 \deleted{known} isolated WDs with $M_{\odot}>1.3$ have $B>10^6$~G~\citep{2004A&A...420..507N}. Recent \textit{Gaia} observations in the solar neighborhood ( $d\lesssim$~100~pc) show the presence of a substantial amount of massive WDs and a bifurcation in the color magnitude diagram that are consistent with a merger formation \citep{2018MNRAS.479L.113K}. \cite{2017A&A...602A..16T} analyzed the sample of WD within 20~pc and show that the amount of WDs in double WD systems ($\sim$25\%) are smaller than the percentage of solar-type main sequence stars in double systems ($\sim$50\%). They conclude that this discrepancy is consistent with about 10-30\% of all isolated WD being the result of a merger. 

The merger of WDs with different cores compositions leads to several final results \citep{2014MNRAS.438...14D}. The merger of two CO WD progenitors results in a near-Chandrasekhar mass product. The remnant consists in a cold core formed by the primary, a hot envelope made by a fraction of the secondary mass, and a disk containing the remaining of the secondary. A small amount of mass, 
about $10^{-3}M_{\odot}$,
is ejected and leaves the system~\citep{2009A&A...500.1193L}. The exact percentage of the secondary mass
in the disk varies according to the mass of the progenitors. According to previous simulations, a good estimate for this percentage is $\sim$~50\% of the less massive progenitor~\citep{2018ApJ...857..134B}, which gives an initial mass for the disk of the order of $10^{-1}$~$M_{\odot}$.

This newborn WD is also expected to have a \replaced{short period}{fast rotation} right after the coalescence. \cite{2018ApJ...857..134B} state that the remnant (cold core and envelope) spins as a rigid body. In contrast, \cite{2007MNRAS.380..933Y} argue that the cold core rotates as a rigid body \replaced{whereas}{and} the envelope spins differentially leading the photosphere to present a quasi-Keplerian angular velocity. This differential rotation, however, vanishes quickly and the remnant eventually starts to rotate uniformly. It is also expected an enhancement of the magnetic field during the coalescence~\citep{2013ApJ...773..136J,2015ApJ...806L...1Z} and in the early years after the merger~\citep{2012ApJ...749...25G}. All those previous properties - presence of the disk, \added{fast} spin\deleted{ period}, and huge magnetic field - are consistent with the observations of \ffu. Thus, if the accreting WD model for \fu is correct, this object probably is a young product of a merger of two less massive CO WDs.

Moreover, the remnant is expected to accrete matter from the disk during its early years, which is also consistent with the proposed model. However, how the disk and the accretion rate evolve is still question for debate. \cite{2013MNRAS.431.2778K} argue that the disk is viscous supported and can be described by the $\alpha$-formalism of \cite{1973A&A....24..337S}. In this model, the accretion rate  would be about\replaced{ $10^{-2}$~$M_{\odot}$\,s$^{-1}$ in the early years}{$10^{-1}$~$M_{\odot}$\,s$^{-1}$ in the first second, rapidly decreasing} \citep{2018ApJ...857..134B}. For those high accretion rates and a rigid-body rotation, the massive WD would \replaced{spun-down}{spin up}
in the first years and would probably break~\citep{2013MNRAS.431.2778K}.
Therefore, we do not \replaced{consider}{take into account} this evolutionary hypothesis. On the other hand, \cite{2007MNRAS.380..933Y} argue that the disk is more likely to be thermal-pressure supported and the early accretion rate is of the order of $10^{-7}M_{\odot}$\,yr$^{-1}$  \citep{2018ApJ...857..134B}. Considering a initial spin period of about 2.5~s~\citep{2018ApJ...857..134B} and an accretion rate smaller than the Eddington limit, we have $R_m>R_c$ in the early years. Thus, the remnant would initially pass through a propeller phase. This propeller phase \replaced{spun-down}{spin down} the remnant, protecting it from breaking. As a consequence of the spin-down, the period and, consequently, the corotational radius increase thus enabling the WD to accrete matter from the disk. 

How this very massive WD would evolve after \replaced{accretion starts}{the start of the accretion} is hard to foresee, once neither the evolution for the post-merger product nor the fate of very massive accreting WDs are well understood. In either case, \fu would be a probable \replaced{candidate to become a NS, by collapse or Supernova~(SN)~Ia}{candidate to become a Supernova~(SN)~Ia or a NS by collapse}. It is also possible \replaced{for}{that} \fu \replaced{to survive}{survives} as an extremely massive WD, such as RE~J0317-853.

\replaced{The origin of SN Ia is still not well known and several models have  been proposed \citep{2014ARA&A..52..107M}. The possibility that AXPs/SGRs as WD pulsars were good candidates for SN Ia were observed by \cite{2014ARA&A..52..107M}. In this case, the rapid rotation delays the explosion \citep{2004ApJ...615..444S,2007MNRAS.380..933Y}, in a similar mechanism to the ``spin-up/spin-down" SN~Ia model. The spin-down effect leads the WD to smoothly decrease the rotation period and, consequently, loses stability. Eventually, the WD carbon-core explodes.}{Several astrophysical objects have been proposed as SN Ia progenitors, as discussed in the review of \citet{2014ARA&A..52..107M}. The AXPs/SGRs as high-mass WDs are cited as good candidates for the SN Ia ``spin-up/spin-down" scenario. In this case,a rapid-rotating high-mass WD, which is rotation supported against ignition, can eventually explode after a period of spin down~\citep{2014ARA&A..52..107M}. This outcome is possible for \fu since we predict a fast spinning-down  Chandrasekhar WD.}

\replaced{For the accreting WD model, both the spin-down and in-falling matter can disturb its stability.
\cite{2004ApJ...615..444S} argue that a WD formed by the merger of two CO WD would not explode as a SN~Ia because it would inevitably become a O-Ne-Mg WD.}{Nonetheless, the accretion rate can also disturb the stability of isolated high mass WDs. For an accretion rate in the range of $10^{-5}$ -- $10^{-6}$~$M_{\odot}$\,yr$^{-1}$, \cite{2004ApJ...615..444S} argue that a WD formed by the merger of two CO WD would not explode as a SN~Ia because it would inevitably become a O-Ne-Mg WD.} In this case, the accretion could lead \replaced{4U~0142+61}{the WD} to exceed the \deleted{WD} limiting mass and become a NS by carbon deflagration collapse \citep{1991ApJ...367L..19N}. \deleted{However, they consider an accretion rate in the range  $10^{-5}$ -- $10^{-6}$~$M_{\odot}$\,yr$^{-1}$ and do not take into account the effect of the magnetic field nor rotation, which are essential ingredients to predict the evolution of WD merger products.}

Conversely, \cite{2007MNRAS.380..933Y} consider\deleted{ed} accretion rates smaller than $10^{-6}$~$M_{\odot}$\,yr$^{-1}$\added{, which are consistent with our accretion rate estimates for \fuu.}\deleted{took into account the rotation.} They found that the remnant of two CO WD can lead to a SN Ia after $\sim 10^5$~yr. Thus, if the accreting WD model were correct, \fu is a good candidate to explode as a SN Ia in a small amount of time. On the other hand, if the fate of \fu is the collapse into a NS or a very massive WD, \fu is a priceless object, since it can give clues how the merger of two CO WDs evolves.

\section{Thoughts about bursts, glitches, and radio emission}
\label{sec:glitches}

\deleted{This paper focuses on the quiescent emission of \ffu. However, an important observational property of \fu is the occurrence of bursts and glitches~}\citep{2011ApJ...736..138G,2014ApJ...784...37D,2017ApJ...834..163A}. \deleted{Below, we discuss qualitatively possible
mechanisms for bursts and glitches in a WD accreting scenario. }

\added{This paper focuses on the quiescent emission of \ffu. However, \fu has bursts and glitches~\citep{2011ApJ...736..138G,2014ApJ...784...37D,2017ApJ...834..163A}, which are nonetheless less energetic compared to the bulk of magnetar bursts \citep{2017ApJ...835...68G}. Rotational instabilities in very fast-spinning WDs can explain these events. This mechanism is particularly appealing for \fuu, which has a period of 8.68~s, not far from the breaking frequency of a WD \citep{malheiro+2012}. }

\replaced{With respect to bursts, the increase in luminosity could}{The increase of luminosity in bursts can}
be generated by thermonuclear runways of carbon~\citep{2005PhRvC..72b5806G}. In fact, thermonuclear reaction are seen in some classes of accreting objects, such as \deleted{in the} NS X-ray busters \citep{1982ApJ...256..637A,1993SSRv...62..223L} and WD recurrent novas \citep{1987ApJ...314..653W,2003cvs..book.....W}.

The glitches in SGR/AXPs \replaced{present}{are} a decrease in the in the spin and in the spin-down. \replaced{Observations show that the glitches happened in the AXPs/SGRs, which can be accompanied by increase in the luminosity}{The glitches in AXPs/SGRs can be accompanied by an increase in the luminosity} \citep{2014ApJ...784...37D}. 
Those glitches \replaced{could}{can} have the same phenomenology proposed  by \cite{1994ApJ...427..984U} and \cite{malheiro+2012} for the WD pulsar model. In this case, a near-Chandrasekhar WD can present sudden
changes in the spin, thereby decreasing the centrifugal forces of the core. Thus, gravity \replaced{would put the WD in a less oblate shape}{would make the WD less oblate}, which would lead to a change in the gravitational energy and consequent release of energy.

\replaced{Regarding the proposed radio emission~(see Sec.~\ref{sec:data}), some clues about its origin can be traced.}
{Some clues about the origin of the \fu  radio emission~(see Sec.~\ref{sec:data}) can be traced.} 
Radio emission is \replaced{relatively common among}{observed in some} accreting WDs. According to \cite{2017AJ....154..252B}\added{,} 21 out of 121 magnetic cataclysmic variable have detected radio emission. The most plausible interpretations are gyrosynchrotron, for the weakly polarized radio emitters, and electron-cyclotron maser emission, for the highly polarized \added{sources}. Since we do not know the polarization of the radio emission of \ffu, both interpretations are possible.

\section{Conclusions}
\label{sec:conclusion}

We \replaced{use}{have used} observational data of  \fu from the mid-IR to the hard X-rays to \replaced{verify the plausibility of a scenario in which the the observed emission is explained by an accreting magnetic WD with a debris disk.}{investigate the scenario of an accreting magnetic WD with a debris disk.} \added{It is essential to mention that we do not claim that this scenario is valid for all SGRs/AXPs.}

We obtain a good fit for the entire SED of \fu using plausible components and parameters. The bremsstrahlung emission from the \replaced{accretion column}{post-shock region} of \fu reproduce\added{s} the hard X-rays. A fraction of this emission is reprocessed by the WD photosphere, originating hot spots\added{,} which respond by the soft X-rays. The bremmsstrahlung emission extends to the optical range, in which \replaced{there are also the contribution from the disk and from the WD photosphere}{the disk and the WD photosphere also contribute}. The disk dominates the emission in the IR range.

The emission from the hard X-rays bremsstrahlung implies a near-Chandrasekhar WD, for which we assume a mass of 1.41~$M_{\odot}$ and a radius of 1,021~km. 
From the fit of the optical/IR emission, we obtain \replaced{an}{a} WD effective temperature of \replaced{$9.5\times10^4$~K}{$9.4\times10^4$~K}.
Those WD characteristics point out to an young WD, having
an estimate age of few~Myr.
The disk has \deleted{an} inner and outer temperatures of 
\replaced{1\,947 and 295~K}{1\,991 and 285~K}\added{, respectively}. \replaced{The modelled disk has temperatures}{Those temperatures are} consistent with
disks seen around WD.

From the spin-down \added{rate}, we can estimate a magnetic field of $\sim 10^7$~G, which is consistent with estimated values in magnetic WDs.

This way, we were able to present a model that explains all the quiescent emission of 4U~0142+61, as well as the observed spin-down. Such a WD can be understood as the result of a recent merger of two less massive WDs. In this scenario, 4U~0142+61 is a good candidate to \replaced{becoming}{become} a SN Ia.

We also estimated a new value for the \fu distance\replaced{ equals to}{,} $3.78^{+0.12}_{-0.18}$~kpc, based on the 3D reddening map of \cite{2018MNRAS.478..651G,2019arXiv190502734G}.

\section*{Acknowledgements}

We thank the anonymous referee\added{s} for \replaced{his/her}{their} comments, which helped us to improve the manuscript.
We acknowledge T. Enoto for kindly provide us with Suzaku soft X-ray data of \ffu.

The authors thank the grant \#2013/26258-4, S\~ao Paulo Research Foundation (FAPESP). CVR acknowledge CNPq (Proc. 303444/2018-5). JGC is likewise grateful to the grant \#2013/15088-0, S\~ao Paulo Research Foundation (FAPESP), and CNPq (Proc. 421265/2018-3 and 305369/2018-0). MC acknowledges financial support under grant \#2015/25972-0 from S\~ao Paulo Research Foundation (FAPESP).

\facility{INTEGRAL}

\software{XSPEC (Arnaud 1996), INTEGRAL Off-line Scientific Analysis package}

\bibliography{biblio}{}
\bibliographystyle{aasjournal}



\listofchanges

\end{document}